\documentstyle[aps,prl]{revtex}
\begin{document}
\draft
\author{Robert de Mello Koch$^{1}$ and Jo\~ao P. Rodrigues$^{2}$,
}
\address{
Department of Physics, Brown University$^{1}$,\\
Providence RI, 02912, USA\\
Department of Physics and Centre for Non Linear Studies$^{2}$,\\ 
University of
the Witwatersrand, Wits 2050, South Africa\\}
\title{
Solving Four Dimensional Field Theories with the Dirichlet Fivebrane.}

\maketitle
\begin{minipage}{\textwidth}
\begin{quotation}
\begin{abstract}
The realization of ${\cal N}=2$ four dimensional super Yang-Mills theories
in terms of a single Dirichlet fivebrane in type IIB string
theory is considered. A classical brane computation reproduces the full
quantum low energy effective action. This result has a simple explanation
in terms of mirror symmetry.
\bigskip
\end{abstract}
\end{quotation}
\end{minipage}

A particularly fruitful approach to the study of supersymmetric quantum
field theories has been to realize these theories as a limit of string or
M theory where gravitational effects decouple. There are two 
complementary approaches to this problem
- the geometric engineering ~\cite{Ge} approach and the Hanany-Witten
brane set up ~\cite{HW}. 
To study ${\cal N}=2$ super Yang-Mills theories in $3+1$
dimensions within the geometric engineering approach, one typically
compactifies type IIA/B string theory on a Calabi-Yau threefold. The full
non-perturbative solution of the ${\cal N}=2$ super Yang-Mills theory is then
obtained by invoking mirror symmetry. In the Hanany-Witten approach, one 
typically considers a web of branes in a flat space. In order to study
${\cal N}=2$ super Yang-Mills theory in $3+1$ dimensions, one considers
two parallel solitonic fivebranes with a number of Dirichlet
fourbranes stretched between them~\cite{EW}. 
In this approach, all perturbative and non-perturbative corrections to the
field theory are coded into the shape of the branes. The solution of these
theories is performed by lifting to M theory. After the lift, the original
type IIA brane set up is reinterpreted as a single fivebrane in M theory,
wrapping the Seiberg-Witten curve $\Sigma$. The relationship between
these two approaches has been explained in ~\cite{Lust}. 
In this report, we will
study ${\cal N}=2$ super Yang-Mills theory using the Hanany-Witten approach.

Up to now, the description of the M theory fivebrane relevant
for ${\cal N}=2$ super Yang-Mills theory has been in terms of eleven 
dimensional supergravity, which is a valid description of M theory at low
energy~\cite{West}. A number of holomorphic quantities ~\cite{Hol}
including the exact low energy effective action ~\cite{West}
can be recovered using the supergravity description.
The supergravity description corresponds to a strong coupling description
of the original type IIA setup. However, one expects the field theory to emerge
in the opposite limit, where the string theory is 
weakly coupled~\cite{HO}. This limit is
not captured by the supergravity approximation, so that one expects that the 
supergravity approach will only be capable of reproducing field theory
quantities which are protected by supersymmetry. 

In this note, we will provide a direct construction in string theory which
realizes the ${\cal N}=2$ super
Yang-Mills theory in terms of a single Dirichlet fivebrane wrapping the
Seiberg-Witten curve.
We will be mainly concerned with two important issues: how a matrix 
description is obtained
and how the string theory configurations described in this article 
are related to the original type IIA brane set up ~\cite{EW}. In particular
the single D5 in Type IIB string theory will be seen to be related
by T-duality to what has been described in the literature as the "magnetic"
IIA brane configuration. We will then show how the D5 
provides a strongly coupled, low energy description of weakly coupled IIA 
string theory in the original brane set up.

We will start with a brane construction consisting of a number 
of Dirichlet fourbranes suspended between Dirichlet sixbranes in type IIA string
theory on $R^{9}\times S^{1}$. The coordinate $x^{7}$ is compact, with radius
$R_{7}$. In the classical approximation, the sixbranes are located at
$x^{8}=x^{9}=0$ and at some fixed $x^{6}$. The world volume
coordinates for the sixbranes are $x^{0},x^{1},x^{2},x^{3},x^{4},x^{5},x^{7}$.
The fourbranes are located at $x^{8}=x^{9}=0$ and at some fixed values of
$x^{4},x^{5},x^{7}$. The fourbranes have world volume coordinates 
$x^{0},x^{1},x^{2},x^{3},x^{6}$. Since the fourbranes stretch between the two
sixbranes, the $x^{6}$ coordinate is restricted to a finite interval. This 
brane configuration is related to the configuration studied in~\cite{Dia} by T
duality along $x^{1},x^{2},x^{3}$. The supersymmetries preserved by the
fourbranes are of the 
form~\cite{Pol} $\epsilon_{L}Q_{L}+\epsilon_{R}Q_{R}$ where
$\epsilon_{L}=\Gamma_{0}\Gamma_{1}\Gamma_{2}\Gamma_{3}\Gamma_{6}\epsilon_{R}$.
Thus the fourbrane breaks one half of the supersymmetry. The sixbranes preserve
supersymmetries of the form $\epsilon_{L}Q_{L}+\epsilon_{R}Q_{R}$ where
$\epsilon_{L}=\Gamma_{0}\Gamma_{1}\Gamma_{2}\Gamma_{3}\Gamma_{4}\Gamma_{5}
\Gamma_{7}\epsilon_{R}$, which breaks half of the remaining supersymmetry.
This leaves a total of ${\cal N}=2$ supersymmetry in $3+1$ dimensions. The
super Yang-Mills theory we wish to study is realized on the world volume of
the fourbrane. The coordinates of the field theory are 
$x^{0},x^{1},x^{2},x^{3}$. 
The essential ingredient allowing a solution of the
field theory, is the realization that by peforming a T duality along $x^{7}$
one obtains a single Dirichlet fivebrane in type IIB string theory. 
This Dirichlet fivebrane has the form $R^{4}\times\Sigma$ where $R^{4}$
is parametrized by the world volume coordinates $x^{0},x^{1},x^{2},x^{3}$
and $\Sigma$ is a surface in the four dimensional space parametrized by the
coordinates $x^{4},x^{5},x^{6},x^{7}$. The requirement that we preserve
${\cal N}=2$ supersymmetry and reproduce the required assymptotic brane geometry
implies that $\Sigma$ is the Seiberg-Witten curve~\cite{EW}. 

We will focus on the simplest case of pure gauge theory. The 
Riemann surface relevant for a configuration of $k$ fourbranes in the original
IIA set up is~\cite{EW}

\begin{eqnarray}
\nonumber
t^{2}&&-2B(v)t+1=0,\qquad
B(v)=v^{k}+u_{2}v^{k-2}+u_{3}v^{k-3}+...+u_{k},\\
t&&=exp(-s/\tilde{R}_{7})=exp(-(x^{6}+ix^{7})/\tilde{R}_{7}),\qquad
v=x^{4}+ix^{5}\qquad \tilde{R}_{7}=l_{s}^{2}/R_{7}
\label{SWCurve}
\end{eqnarray}

\noindent
This curve corresponds to a fivebrane with two assymptotic sheets connected
by $k$ tubes. The two assymptotic sheets are T dual to the sixbranes in the
above IIA setup, whilst the tubes are T dual to the fourbranes.
Our first task is to provide a classical description of a fivebrane 
with this geometry. Since we are interested in describing
the world volume of the fourbranes in the original IIA description, it is
most natural to use the world volume co-ordinates
$x^{0},x^{1},x^{2},x^{3},x^{6},x^{7}$. This is different from the approach 
followed in ~\cite{West}. 

Consider
the low energy description of the Dirichlet fivebrane, which is known to be a
$5+1$ dimensional super Yang-Mills theory~\cite{EdW}. 
The bosonic part of the fivebrane Lagrangian is

\begin{equation}
{\cal L}=
Tr\Big(F_{\mu\nu}F^{\mu\nu}+D_{\mu}X^{I}D^{\mu}X^{I}+\big[
X^{I},X^{J}\big]^{2}\Big),
\label{BosonicPart}
\end{equation}

\noindent
where $I=4,5,8,9$ and $\mu,\nu =0,1,2,3,6,7$. The $X^{I}$ are $k\times k$
dimensional matrices.

The classical configuration corresponding to a 
Dirichlet fivebrane wrapped on the
Seiberg-Witten curve is $X^{8}=X^{9}=0$ with $X^{4}$ and $X^{5}$ simultaneously
diagonal. The eigenvalues $x_{i}^{4}$ and $x_{i}^{5}$ 
of $X^{4}$ and $X^{5}$ depend on $x^{6}$ and $x^{7}$ as
we now explain. Once a value for $x^{6}$ and $x^{7}$ is given,~(\ref{SWCurve})
may be solved for the $k$ roots $v_{i}$. The real part of $v_{i}$ then gives
$x_{i}^{4},$ whilst the imaginary part gives $x_{i}^{5}$. Clearly, once we
specify a point on the membrane world volume, $x_{i}^{4}$ and $x_{i}^{5}$ range over
the correct coordinates to be identified either with a given tube connecting the two
parallel sheets of the fivebrane, or with one of the $k$ "circular" arcs 
on these parallel sheets themselves. 
The $k^{2}-k$ off diagonal entries in 
$X^{4}$ and $X^{5}$ describe the fivebrane self interaction arising from open 
strings stretching between these tubes. These off diagonal entries, as well as 
the gauge field, are set
to zero in the classical configuration which we study.

We remark that the above fivebrane solution is obtained by identifying
the eigenvalues of the fivebrane's matrix co-ordinates with the zeroes (in
$v$) of the Seiberg-Witten curve. This is familiar from the collective field
approach to the large $N$ limit of matrix models, where the collective field
describing the density of eigenvalues can be identified with the density of
zeros of a suitable polynomial~\cite{us}. 

We now show how the exact Seiberg-Witten~\cite{SW} 
low energy effective action is reproduced. The terms in the
matrix theory Lagrangian giving rise to the scalar kinetic term of the four 
dimensional field theory are $(m=0,1,2,3)$

\begin{equation}
{\cal L}_{kin}=\int d^{2}s Tr\Big(\partial_{m}Y\partial^{m}
Y^{\dagger}\Big)=\int d^{2}s \partial_{m}y_{i}\partial^{m}\bar{y}_{i},
\label{RelTerms}
\end{equation}

\noindent
where $y_{i}$ are the diagonal elements of $Y=X^{4} + i X^{5}$. Later we will use the
fact that $y_{i}$ are simply the roots $v(t)$ described by equation
~(\ref{SWCurve})\footnote{We don't worry about dimensions in the present discussion. This will
be discussed in a later section. We have also set $\Lambda = 1$}.
For concreteness, we now consider $N_{c}=2$ in which case we have
($y_{1}^{2}=y_{2}^{2}=v^{2}\equiv y^{2}$)

\begin{equation}
y^{2}=-u+\cosh (s/\tilde{R}_{7}).
\label{Explicit}
\end{equation}

\noindent
To evaluate the integral in~(\ref{RelTerms}) 
we can proceed as follows: Since the only
$x^{m}$ dependence in $y$ is contained in $u$, we may rewrite
~(\ref{RelTerms})  as

\begin{equation}
{\cal L}_{kin}=\partial_{m}u\partial^{m}\bar{u}\int_{\Sigma} 
\lambda\wedge\bar{\lambda}\quad , \qquad \quad
\lambda={\partial\over\partial u}\Big[
\sqrt{\cosh({s\over \tilde{R}_{7}})-u}\Big]ds
\equiv {\partial\tilde{\lambda}\over\partial u}.
\label{RlTerms}
\end{equation}

\noindent
Note that $\tilde{\lambda}$ is a meromorphic one form, with a double pole at
$t=\infty$, and that $\lambda$ is a holomorphic one form. Choose a
symplectic basis of the first homology class of $\Sigma$ denoted
$\alpha,\beta$. Now, using the Riemann bilinear identity for abelian
differentials of the first kind~\cite{Spring}, 
~(\ref{RelTerms})  can be expressed as

\begin{eqnarray}
\nonumber
{\cal L}_{kin}&&={1\over 2i}\Big(\partial_{m}a\partial^{m}\bar{a}_{D}
-\partial_{m}\bar{a}\partial^{m}a_{D}\Big)=Im(\partial_{m}
a\partial^{m}\bar{a}_{D}), \\
a&&=\oint_{\alpha}{ds\over \tilde{R}_{7}}
\sqrt{\cosh({s\over \tilde{R}_{7}})-u}=
\oint_{\alpha}dx{\sqrt{x-u}\over \sqrt{x^{2}-1}}\quad
a_{D}=\oint_{\beta}{ds\over \tilde{R}_{7}}
\sqrt{\cosh({s\over \tilde{R}_{7}})-u}=
\oint_{\beta}dx{\sqrt{x-u}\over \sqrt{x^{2}-1}}.
\label{Expressions}
\end{eqnarray}

\noindent
The classical brane solution has reproduced the correct
scalar kinetic terms of the exact low energy effective action. 
$\tilde{\lambda}$ is in fact the
Seiberg Witten differential, remarkably in its original form~\cite{SW}. 
In view of the comments following~(\ref{RelTerms}) 
it is clear that our approach gives the Seiberg Witten differential,
in the general case, as $\lambda_{SW}=v(t){dt/t}$.
This agrees nicely with the known results ~\cite{Known}.

The relationship between our brane setup and the 
one used in~\cite{EW}\footnote{Related ideas were 
considered in~\cite{Guk},~\cite{Kap}.}
is most easily demonstrated by considering
the ${\cal N}=2$ super Yang-Mills theory on $R^{3}\times S^{1}$. 
Consider the system described earlier of parallel Dirichlet sixbranes 
with $N_{c}$ Dirichlet fourbranes
stretching between them. We take $x^{1}$ to be compact with radius $R_{1}$,
$x^{7}$ to be compact with radius $R_{7}$ and $g_{s}=R_{11}l_{s}^{-1}$
\footnote{$R_{11}$ should be thought of as a parameter, which we could have also
chosen as $g_{s}$. The lift to M theory is described later.}. The
sixbranes wrap the compact $x^{1}$ and $x^{7}$ directions; the fourbranes
wrap $x^{1}$. Performing a T duality along the cycle of length $R_{7}$, one
obtains a single Dirichlet fivebrane as discussed above. This Dirichlet
fivebrane wraps the Seiberg-Witten curve $\Sigma$ and the compact $x^{1}$
direction and has string coupling  
$g_{s}={R_{11}\over R_{7}}$. Now perform a T duality along the cycle of length
$R_{1}$. We obtain a single Dirichlet fourbrane in type IIA string theory,
wrapped on the Seiberg Witten curve $\Sigma$. After this second T duality,
$x^{1}$ is compact with radius ${l_{s}^{2}\over R_{1}},$ $x^{7}$ is compact
with radius ${l_{s}^{2}\over R_{7}}$ and $g_{s}={R_{11}l_{s}\over R_{1}R_{7}}$.
Lifting to M theory, we obtain a single M theory fivebrane wrapping the 
Seiberg Witten curve $\Sigma$ and $x^{11}$. The $x^{1}$ direction is compact
with radius $\sqrt{R_{7}l_{p}^{3}\over R_{1}R_{11}}$, the $x^{7}$ direction is
compact with radius $\sqrt{R_{1}l_{p}^{3}\over R_{7}R_{11}}$ and $x^{11}$ is
compact with radius $\sqrt{R_{11}l_{p}^{3}\over R_{1}R_{7}}.$
Using the eleven dimensional
Lorentz invariance of M theory, we may reinterpret $x^{7}$ as the dimension
which grows at strong coupling. Since the Planck length of the theory is held
fixed, the string tension is transformed as

\begin{equation}
l^{2}_{s}\to l^{'2}_{s}=(R_{11}l^{2}_{s})/R_{1},
\label{NewTension}
\end{equation}

\noindent
as explained in~\cite{JPR}. In this case, we obtain two parallel
solitonic fivebranes in IIA string theory, with $N_{c}$ Dirichlet fourbranes
stretched between them. The string coupling is 
$g_{s}'=(R_{1}l_{s}')/(R_{11}R_{7}).$ The
solitonic fivebranes and the Dirichlet fourbranes both wrap the $x^{11}$
direction which has a radius $l_{s}^{'2}/ R_{7}$. 

This compactification 
has been considered in~\cite{Kap}\footnote{Note however that the coordinate $x^{6}$
is not compact in our case, i.e., we're not considering the elliptic case.}
where it was argued that the link between
the two solitonic fivebranes with $N_{c}$ fourbranes (the "electric" IIA brane
configuration) and the single Dirichlet fourbrane wrapping the Seiberg-Witten 
curve (the "magnetic" IIA brane configuration) is in fact a mirror transform. 
In the present context, this can independently be seen as follows:
under T duality along $x^{11}$,
the electric IIA brane configuration is mapped into two solitonic
fivebranes with $N_{c}$ Dirichlet threebranes sretched between them. Our 
starting brane configuration, consisting of $N_{c}$ Dirichlet fourbranes 
stretched between two Dirichlet sixbranes
is mapped into two Dirichlet fivebranes with
$N_{c}$ Dirichlet threebranes stretched between them under T duality
along $x^{1}$. According to~\cite{HW}, these
IIB brane configurations are related by mirror symmetry. 
The mirror transform maps the Coulomb branch of the electric theory to the
Higgs branch of the magnetic theory~\cite{SI},~\cite{HW}, 
which does not receive string loop corrections~\cite{SI} so that the 
classical calculation is exact. This provides a simple explanation for why the
classical Dirichlet fivebrane is capable of reproducing the full quantum 
effective action of the ${\cal N}=2$ gauge theory.

The electrical BPS
state, which is a fundamental string stretching between two
Dirichlet fourbranes in the electric IIA brane configuration,
becomes in the single Dirichlet 
fivebrane description,
a Dirichlet three brane
wrapping $x^{1}$ (with radius $R_{1}$), $x^{7}$ (with radius $\tilde{R}_7$)
and stretching between the two tubes of the fivebrane. Its mass is given
by
\footnote{Further 
details can be found in ~\cite{prep}.}

\begin{equation}
m=2\pi R_{1}\times 2\pi\tilde{R}_{7}\times 
{2 |a(u)|\over \alpha}\times {1\over g_{s}l_{s}^{4}}
= {R_1 \over R_{11} l_s^{2}}{8 \pi^2 |a(u)|\over \alpha}
= {1 \over l_s^{'2}}{8 \pi^2 |a(u)|\over \alpha}.
\label{Mass}
\end{equation}

\noindent
We have let $y={1 \over \alpha} v$, with $\alpha$ a parameter 
with the dimensions of 
$L^{-2}$, so that $y$ is a displacement and
$v$ has the usual dimensions of a Higgs field. We have also
used $g_s=R_{11}/R_7$ for the D5 string coupling 
and equation~(\ref{NewTension}). The magnetic
BPS state, which is a Dirichlet two brane in the 
electric IIA brane setup stretching across the hole between the two 
Dirichlet fourbranes and the two solitonic fivebranes,
becomes a Dirichlet threebrane in the single D5 description 
has mass ~\cite{prep}:

\begin{equation}
m=2\pi R_{1}\times \tilde{R}_{7}\times {|a_{D}(u)| \over \alpha}\times
{1\over gl_{s}^{4}}
= {R_1 \over R_{11} l_s^{2}}{2 \pi |a_{D}(u)|\over \alpha}
= {1 \over g'_s l_s^{'3}}{2 \pi \tilde{R}_7 |a(u)|\over \alpha},
\label{LKJLK}
\end{equation}

\noindent
where $g'_s$ is the string coupling constant in the D4/NS5 setup. 

We now discuss the region of applicability of the D5 picture. ${\cal N}=2$
super Yang-Mills has a characteristic mass scale $\Lambda$, and therefore one
way to approach the problem is to consider scaling limits 
$\Lambda R_1 = \epsilon^{\alpha_1}, ...$ in the limit as $\epsilon \to 0$. 
There are four parameters $R_1$, $R_{11}$, $\tilde{R}_7$ and $l_s$. One constraint
amongst them is that the size of the $x^{11}$ coordinate in the original D4/NS5
set up is given by 

\begin{equation}
r_{11} = l_{s}^{'2} / R_{7} = (R_{11} \tilde{R}_7) / R_1 . 
\label{Size}
\end{equation}   

If we wish to describe ${\cal N}=2$
super Yang-Mills on $R^4$, we may then choose $r_{11}$ to grow in a specified manner.
Other possible constraints depend crucially on how lengths are obtained from
the field theory Higgs (i.e., the parameter $\alpha$ in equations~(\ref{Mass})
and~(\ref{LKJLK}))
and a possible specification relating field theoretic masses to D5/M5 descriptions. 
Clearly, there is a preferred choice for $\alpha$: $\alpha = l_{s}^{'-2}$, where $l_{s}'$ is the
string coupling constant in the D4/NS5 set up. For this choice, the BPS masses 
~(\ref{Mass}) and~(\ref{LKJLK}) are 
immediatly field theoretic masses. For this choice of $\alpha$
there are no further constraints amongst the parameters.
 
We then have for the single D5 string coupling constant
$g_s$

\begin{equation}
g_s = (R_{11}\tilde{R}_7) /l_s^2 = (r_{11} R_1) / l_s^2 ,    
\label{StringCoup}
\end{equation}

\noindent
where we used~(\ref{Size}) in the 
second equality above. We require $l_s \to 0$ so that the 
Yang-Mills description of the D5 is valid, $r_{11}$ and $R_1$ to be large;
as a result the D5 will be strongly coupled. This is to be expected since $r_{11}$, which is
the size of the world volume coordinate $x^{11}$ in the D4/NS5 picture becomes the 
M-theory circle on the magnetic side, and therefore decompactification in the
Coulomb phase will in general correspond to strongly coupled magnetic descriptions.

Finally note that the second equality in~(\ref{StringCoup}) 
is independent of $\tilde{R}_7$
and $R_{11}$,
and therefore $\tilde{R}_7$ can be made small by suitably adjusting $R_{11}$. Since
$\tilde{R}_7$ is the M-theory circle in the electric picture, the type IIA string
theory in the D4/NS5 set up is weakly coupled. We also remark that 
the above fivebrane analysis provides a 
fivebrane worldvolume description of
the type IIB setup considered in the first of~\cite{Ge}.


{\it Acknowledgements} 
We would like to thank Mihail Mihailescu, Antal Jevicki, Sanjaye Ramgoolam,
Radu Tatar and especially Jo\~ao  Nunes for helpful discussions.
The work of R.dMK is supported by a South African FRD bursary. The work of 
J.P.R. is partially supported by the FRD under grant number GUN-2034479.


\end{document}